\documentclass[epsf]{emulateapj}
\shorttitle{Internal Dynamics of M31 Globular Clusters}
\shortauthors{Strader et al.}
\def\etal{{\it et al.}}
\def\kms{\,km~s$^{-1}$}
\def\us{\char`\_}
\def\gsim{\;\rlap{\lower 2.5pt
 \hbox{$\sim$}}\raise 1.5pt\hbox{$>$}\;}
\def\lsim{\;\rlap{\lower 2.5pt
   \hbox{$\sim$}}\raise 1.5pt\hbox{$<$}\;}
\usepackage{subfigure}

\begin{document}

\title{Star Clusters in M31: V. Internal Dynamical Trends: Some Troublesome, Some Reassuring}

\author{Jay Strader, Nelson Caldwell, Anil C.~Seth\altaffilmark{1}}
\email{jstrader@cfa.harvard.edu}

\altaffiltext{1}{Harvard-Smithsonian Center for Astrophysics, Cambridge, MA 02138}

\begin{abstract}

We present internal velocity dispersions and precise radial velocities for 200 globular clusters (GCs) in M31 that are derived using new high-resolution spectra from MMT/Hectochelle. Of these, 163 also have King model structural parameters that allow us to estimate their mass-to-light ratios. This is, by far, the largest such dataset available for any galaxy, including the Milky Way. These data strongly confirm earlier suggestions that the optical and near-infrared mass-to-light ratios of M31 GCs decline with increasing metallicity. This behavior is the opposite of that predicted by stellar population models for a standard initial mass function. We show that this phenomenon does not appear to be caused by standard dynamical evolution. A shallower mass function for metal-rich GCs (with $dN/dM \propto M^{-0.8} - M^{-1.3}$ below 1 $M_{\odot}$) can explain the bulk of extant observations. We also observe a consistent, monotonic correlation between mass-to-light ratio and cluster mass. This correlation, in contrast to the correlation with metallicity, is well-explained by the accepted model of dynamical evolution of GCs through mass segregation and the preferential loss of low-mass stars, and these data are among the best available to constrain this process.

\end{abstract}


\section{Introduction}

A fundamental result of stellar population research in recent years is the realization that massive star clusters (typified by old globular clusters---GCs) are not simple stellar populations with a single age and chemical composition. Self-enrichment in light elements among GCs appears ubiquitous (Gratton \etal~2004), perhaps even leading to multiple main sequences in some clusters (Piotto \etal~2007). Further afield, some young massive clusters in the Magellanic Clouds have internal age spreads of several hundred Myr (Mackey \etal~2008), and the most massive extragalactic GCs may self-enrich in iron (Strader \& Smith 2008; Bailin \& Harris 2009).

Increasing evidence from dynamical studies suggests that many GCs deviate from simple stellar populations in another respect: their mass-to-light ratios ($M/L$). Kruijssen \& Mieske (2009) show that a sample of Galactic metal-poor GCs had, on average, $M/L_V$ $\sim 20$\% lower than expected. This difference can be explained by standard dynamical evolution---the preferential depletion of low-mass stars through two-body relaxation and evaporation. However, Strader \etal~(2009) showed that metal-rich M31 GCs (those with [Fe/H] $> -1$) had $M/L_V$ much lower, by a factor of two, than predicted by stellar population models with a standard Kroupa initial mass function (IMF). While there are minimal comparable data for Galactic GCs, the existing observations are consistent with the same result (see data in McLaughlin \& van der Marel 2005). This discrepancy could be principally due either to unexpectedly strong dynamical evolution or significant issues with stellar population synthesis models, including the assumed IMF.

The sample of GCs in Strader \etal~(2009) was relatively small, with only $\sim 14$ metal-rich clusters. In this paper we take advantage of the large GC population of M31 and the wide-field high-resolution Hectochelle spectrograph on the MMT to present a new study of $M/L$ among M31 GCs. This new sample is much larger than that of any previous study, and more than doubles the total number of $M/L$ estimates for GCs in the Local Group. It is also more diverse, covering a wide range of cluster masses and metallicities. In addition, we derive both optical and near-infrared $M/L$ values, which provide better constraints on the stellar populations of these GCs.

This paper is the fifth in a series presenting the results of low and high-resolution spectroscopy of M31 star clusters with the MMT instruments Hectospec and Hectochelle. Earlier papers presented an overview of the survey and a study of young clusters (Caldwell \etal~2009), metallicities and ages of old globular clusters (Caldwell \etal~2011), kinematics of central metal-rich GCs (Morrison \etal~2011), and CN abundances (Schiavon \etal~2011, in preparation). Coordinates for all GCs may be found in Caldwell \etal~(2009), whose naming conventions we follow. Throughout the paper, we adopt a distance of 780 kpc to M31. 

\section{Data Acquisition and Reduction}

All data were acquired with Hectochelle on the MMT. Hectochelle is a multi-object single order spectrograph with resolution $\sim 34000$. Each order is $\sim 150$ \AA\ wide. The instrument is fed by the fiber positioner for Hectospec (Fabricant et al.~2005), with 240 1.5\arcsec\ diameter fibers available for assignment to objects or sky. Because of a planned stellar population study, we used four different orders to cover a number of relevant spectral features (although many GCs were only observed in a subset of these orders). The orders and their central wavelengths are: OB37 (4340\AA), RV31 (5215\AA), OB25 (6540\AA), and Ca19 (8542\AA). The last of these encompasses the \ion{Ca}{2} triplet region, and can be profitably observed even in bright conditions.

We drew our sample from the GC catalogs presented in Caldwell \etal~(2009; 2011), which are revisions and refinements of the Bologna group catalog (Galleti \etal~2006). The field of view of the fiber positioner is $1^{\circ}$ (14 kpc at the distance of M31). We used two different fields: one centered on the nucleus, and another centered $0.5^{\circ}$ along the major axis to the SW. Our sample therefore includes objects out to a projected radius of $\sim 12$ kpc, including many metal-poor GCs, although we miss objects in the outer halo. A list of the pointings, including the orders observed, is given in Table 1. Exposures were between two and four hours per field. In total, we observed 285 clusters; 213 have formal ages from low-resolution spectroscopy $> 4$ Gyr (Caldwell \etal~2011); 63 have younger ages, with the remaining 9 of unknown age. 

The spectra were extracted from the CCD images in the manner described in Caldwell et al.~(2009). Before sky subtraction, it was necessary to normalize the spectra to account for variations in the fiber throughput. This correction was estimated using the strength of several night sky emission lines in the appropriate order. Sky subtraction was performed using the average of 20--30 sky fibers. We discuss the determination of radial velocities and velocity dispersions in \S 3. Our limiting magnitudes to estimate radial velocities and velocity dispersions (for old GCs) are $V \sim 19.5$ and $18.5$, respectively.

\section{Analysis}

\subsection{Radial Velocities}

We derived heliocentric radial velocities through cross-correlation with synthetic or observed template spectra. For data taken in the RV31 order, we used a synthetic stellar spectrum of a dwarf solar abundance star with an effective temperature of 5500 K. We used contemporaneous observations of G and K giant velocity standards for the other orders. We used the SAO routine \emph{xcsao} for the cross-correlation. Final cluster velocities were drawn from the star that gave the highest cross-correlation R value (a statistic measuring the relative signal-to-noise of the cross-correlation peak). To be taken as a reliable velocity, this R value was required to be higher than 4 for the RV31 order, 5 for Ca19, and 6 for OB25 and OB37;  these limits were determined by comparing the Hectochelle radial velocities with Hectospec velocities reported by Caldwell \etal~(2011). We note that the resolution of the Hectochelle data allows the determination of radial velocities even against the high surface brightness of the inner bulge of M31. There are four GCs for which the previous Hectospec data was contaminated by background light (see Caldwell \etal~2011); we can report reliable Hectochelle radial velocities for these GCs.\footnote{The GCs: B070-G133, B132, B262, NB21.}

Repeat measurements in multiple orders were used to estimate the velocity uncertainties. For objects with data in more than one order, the final velocities are a weighted average of the different orders. The uncertainties of these final velocities include a common term of 0.4 \kms\ to account for errors in the wavelength calibration and the zeropoint of the velocity scale. We successfully derived radial velocities for 246 clusters. Of these, 207 are old GCs, 1 of intermediate age, 34 younger objects, and 4 of unknown age. The final radial velocities are listed in Table 2. 

\subsection{Velocity Dispersions}

For the old GCs, we derived integrated velocity dispersions using the method discussed in Strader \etal~(2009). Briefly, after Fourier filtering, the object spectra of M31 GCs were cross-correlated with a set of template spectra of type G to early M. We determined the relationship between the full width at half maximum of the cross-correlation peak and the observed (projected) velocity dispersion $\sigma_{p}$ for each pair of templates using a grid of convolved spectra and interpolation. The median and standard deviation of the individual pair estimates are taken as the best estimate and error of $\sigma_{p}$ for each order. These values are listed in Table 3. Uncertainties in $\sigma_{p}$ are discussed in detail below. The lowest estimates of $\sigma_{p}$ should be considered the most uncertain, and it is possible we have overestimated $\sigma_{p}$ for a few of these low-luminosity clusters (the bias is toward overestimation, since objects scattered low may have unmeasureable $\sigma$ and so not be included.) We did not derive velocity dispersions for the younger clusters, since the measurements for such objects are dependent on accurate stellar template matches.

The velocity dispersions in Table 3 are measured through the finite Hectochelle fiber apertures (radii 0.75\arcsec = 2.84pc). This value is comparable to the half-light radius for most M31 GCs, and therefore these dispersions are intermediate between the central and global values $\sigma_{0}$ and $\sigma_{\infty}$. For typical cluster structural parameters, the seeing has a negligible effect on the inferred velocity dispersion for good seeing. In poor conditions ($> 1$\arcsec), the effect is still small, but potentially noticeable. A subset of our RV31 data (that from 2008) was taken in poor seeing of 1.7\arcsec. In these conditions, for a typical GC with $c=1.6$\footnote{Here we take the concentration $c$ as log ($r_t/r_0$), with tidal radius $r_t$ and King radius $r_0$.} and half-mass radius of 3 pc, $\sigma_{p}$ will be lower by $\sim 3$\% than in good seeing. We use this mean correction for computing masses using the poor-seeing 2008 RV31 data but make no seeing corrections for any other order. The velocity dispersions listed in Table 3 include this seeing correction.

Most of the GCs have useful data in only 1--2 orders, commonly RV31 and OB37. Strader \etal~(2009) demonstrate that there is more variation in estimates of $\sigma_{p}$ among orders than among templates. After accounting for the seeing as described previously, there is still a systematic difference between the $\sigma_p$ values from the 2008 RV31 (Mg$b$) and OB37 (centered at 4350 \AA) data, such that the former are smaller by 3.5\%. Only a small number of GCs in our sample have literature $\sigma_{0}$ measurements derived from multi-order data that we might plausibly use as reference values to assess whether the RV31 or OB37 values are more reliable. We compared the derived $\sigma_{0}$ values (\S 3.3) for these overlapping thirteen GCs with the published estimates taken from the compilation of Strader \etal~(2009). The RV31 values have a median difference of only 2\% compared to the literature values; the median difference for OB37 is 8\%. Therefore, we scale the OB37 values down by the median offset of 3.5\% to put the orders on a common system; this correction is included in the OB37 values in Table 3.

Correcting for this order-dependent offset, the distribution of the differences between RV31 and OB37 is consistent with the nominal $\sigma_{p}$ errors as originally estimated, if an additional uncertainty of 0.4 \kms\ is added in quadrature to each error estimate. Thus, the individual order uncertainties in Table 3 have this additional error included. This procedure also gives generally consistent results for the repeats in RV31 data between 2008 and 2009, although there were only fourteen GCs in common between these setups.

We use the weighted average of the individual order values as our best estimate of $\sigma_{p}$ for each GC. Given the various corrections discussed in this section, we add an uncertainty of 5\% in quadrature to the final weighted $\sigma_{p}$ estimates.

While we took Ca19 (Ca triplet) data for a large sample of GCs, we do not use the Ca19 data for objects which also have RV31 or OB37 $\sigma_p$. This is because the estimates from Ca19 do not agree with the other measurements---they are systematically larger by approximately 10\% for typical GCs (the difference is smaller at higher values of $\sigma_p$). The offset persists for a wide range of Fourier filtering parameters. Our hypothesis is that the offset is caused by the strong damping wings of the Ca triplet, which evidently differ between templates and the integrated light of GCs. 

Fifteen GCs in our sample have only data taken in Ca19. To estimate $\sigma_p$ for these GCs, we performed a linear fit between the Ca19 and OB37 $\sigma_p$ values for a large set of GCs that had both measurements. The resulting relation is $\sigma_{OB} = (1.09\pm0.03) \sigma _{CaT}  - (2.3\pm0.5)$; the 1$\sigma$ scatter in the relation is 1.0 \kms. The ``updated" Ca19 values (which are those given in Table 3), should then be, at least to first order, usable to estimate cluster masses. 

\subsection{Masses and Mass-to-Light Ratios}

We derive GC masses as described in Strader et al.~(2009). Central and global velocity dispersions ($\sigma_{0}$ and $\sigma_{\infty}$) are estimated using $\sigma_{p}$ by integrating a known King model over the fiber aperture. We derive both virial and King model masses; these are listed in Table 4 with the other derived quantities. Virial masses are nominal estimates of the ``global" mass of the system (using the half-mass radius and global velocity dispersion $\sigma_{\infty}$), while King masses utilize the core parameters $r_{0}$ (the King radius, similar but not identical to the core radius $r_c$) and the central velocity dispersion $\sigma_{0}$. Unlike in some studies of resolved systems, we do not have independent estimates of $\sigma_{0}$ and $\sigma_{\infty}$, and so these two mass estimates are not independent; the virial masses are, in the median, 10\% higher than the King masses. Since the virial masses are less sensitive to the accuracy of the King model fit, we adopt them for the remainder of the paper. The uncertainties listed for these masses include the uncertainties in both the structural parameters and velocity dispersions.

We use three principal sources for King model structural parameters. The first is Barmby \etal~(2007), who provide fits in $V$-equivalent bands for 41 GCs in our sample using Advanced Camera for Surveys (ACS), Wide Field and Planetary Camera 2 (WFPC2), or Space Telescope Imaging Spectrograph (STIS) imaging.\footnote{We used the F814W model fit for B082-G144, as the F606W profile was saturated in the core. Barmby \etal~(2009), a paper focused on young clusters, also provides fits for a few old GCs, including B083-G146 in our sample. We use the F450W fit for this object.} The second source is Peacock \etal~(2010; see also Peacock \etal~2009), who give ground-based $K$-band King model fits for a large sample of M31 GCs. Finally, we provide new fits for 73 GCs on the basis of both archival HST imaging and new HST/ACS data through the PHAT survey (Dalcanton et al. 2011, in preparation). The former data are in a wide range of filters, ranging from $B$ to $I$-equivalent, while the latter are principally in F475W. These two-dimensional circular King model fits were performed as described in Larsen \etal~(2002) and Strader \etal~(2009). While we give the model parameters in Table 5, they should be considered preliminary; these fits will be superseded by a comprehensive study of the structural parameters of GCs in M31 once the PHAT survey is complete. We use the PHAT F336W data for 11 GCs that have no other data available. While these compiled structural parameters are derived from a variety of filters, those in the $B$ to $V$ range are most common. These wavelengths are well-matched to the orders in which our velocity dispersions are derived (principally at central wavelengths of 4340\AA\ and 5215 \AA), so our mass estimates should be consistent in this regard.

The Peacock \etal~(2010) structural parameters do not have reported errors. These can be estimated by comparing their concentrations and half-light radii to those of Barmby \etal~(2007). In agreement with the discussion in Peacock \etal~(2009), the structural parameters appear to be unreliable for GCs with total $K$ magnitudes $> 15$, and we do not use the fits below this limit. Brighter than $K=15$, there is no strong correlation between the luminosity of the cluster and the apparent uncertainties in the structural parameters; we assign a common error of 0.15 in $c$ and 0.3 pc in $r_h$ to all of the Peacock \etal~(2010) measurements with $K < 15$. Similarly, because of the manner in which the new HST King model fits were done, it is difficult to assign them self-consistent errors; we adopt fixed percentage errors of 10\% in the size parameters and 15\% in $r_t/r_0$ (so $\sim 0.06$ in $c$), consistent with typical uncertainties for the Barmby \etal~(2007) HST fits.

We calculate $V$ and $K$-band luminosities as follows. For objects with structural parameters from Barmby \etal~(2007), we adopt their integrated $V$ model luminosities. For all other GCs, we use $V$ magnitudes from Caldwell \etal~(2009). We use the $K$ magnitudes of Peacock \etal~(2010) preferentially where available. For the remainder of the GCs, we use the $K$ magnitudes published by Galetti \etal~(2004) from 2MASS.\footnote{A comparison of common clusters suggests that the $K$ magnitude for B196-G246 in Peacock \etal~(2010) is in error, as is the Galetti \etal~(2004) value for B090.} The $K$ magnitudes from Galetti \etal~do not have any published errors. We assign approximate errors to these values using a standard exponential error distribution, with the parameters estimated using the distribution of differences (as a function of magnitude) between the Peacock \etal~(2010) and Galetti \etal~(2004) $K$ magnitudes for nearly 100 GCs in common.

The Barmby \etal~(2007) luminosities assume their values of $E(B-V)$. For nearly all other GCs, we use the reddenings from Caldwell \etal~(2009).\footnote{For B335-V013 we used the values from Fan \etal~(2008). The metallicity listed for this GC is from the same source. For B037-V327 the $E(B-V)$ and $L_V$ are from Strader \etal~(2009).} Three GCs (B056D; B088D; B515) had no published estimates of $E(B-V)$; we assume the median value of $E(B-V) = 0.16$ for these clusters. The resulting estimates of $M/L_{V}$ and $M/L_{K}$ are given in Table 4. As in Strader \etal~(2009), the listed errors for these quantities include an additional 5\% to account for uncertainties in distance and reddening. Table 4 also includes the corresponding spectroscopic estimates of [Fe/H] from Caldwell \etal~(2011) for nearly all of the sample GCs.

In total, new $M/L_{V}$ values are given for 163 M31 GCs. In Table 6, using the same format as Table 4, we update the compilation of M31 GC literature data first presented in Strader \etal~(2009). A larger sample of 30 GCs is included, since some GCs with previously published estimates of $\sigma$ now also have structural parameters measured. The velocity dispersions in this table are weighted averages of the various literature values, and have been updated to include new data where appropriate. For the remainder of the paper, we use a combination of Tables 4 and 6, giving preference to the more robust multi-order literature data for GCs in common with our new sample. There are a total of 178 M31 GCs in the final combined sample, although a subset of these GCs do not have $K$-band luminosities or spectroscopic metallicities. This sample is a factor of six larger than in any previous study of $M/L$ for M31 GCs, and more than doubles the total number of $M/L$ estimates for all GCs in the Local Group.

For clarity, and given our large sample, in subsequent sections we consider only those objects with errors in $M/L_{V}$ that are less than 25\%; this gives total samples of 131 and 126 GCs in $V$ and $K$ respectively. However, all GCs are listed in the relevant tables.

\section{Results}
\subsection{A Troublesome Metallicity Trend}

The principal result of Strader \etal~(2009) was that the $M/L_V$ for old M31 GCs has a shallow, monotonic decline with increasing metallicity. This is opposite basic expectations from stellar population models, which predict fainter optical luminosities (and thus an increase in $M/L$) for more metal-rich systems due to increased line blanketing. Figure 1 strongly confirms this result, but with a much larger sample of GCs (in this and subsequent figures, the point size is inversely proportional to the error in the $M/L$ measurement). We have plotted flexible stellar population synthesis (FSPS) models, assuming an age of 12.5 Gyr and a Kroupa IMF, from Conroy \etal~(2009; updated in Conroy \& Gunn 2010). These models use Padova isochrones (Marigo \etal~2008) and the BaSeL stellar library.\footnote{A subtlety is that these models are constructed with solar-scaled abundances while the GCs are likely to be $\alpha$-enhanced (Schiavon \etal~2011, in preparation); [Fe/H] is plotted for both models and data, and the effects of [$\alpha$/Fe] are certainly subdominant to the other uncertainties in the comparison.} We have assumed a fraction of 50\% blue horizontal branch stars for [Fe/H] $< -1$ and 10\% for higher metallicities; the exact values used have very little effect on the predicted $M/L_V$.\footnote{For example, at typical metallicities of $-1.4$ and $-0.4$, the maximum variation caused by assigning all stars to a blue horizontal branch is $\sim 3$\% in $M/L_V$ and $\sim 5$\% in $M/L_K$ (using the prescription in Conroy \etal~2009: stars on the nominal blue horizontal branch are evenly distributed between the red clump and $\sim 16000$ K.)}

\begin{figure*}
\epsscale{0.7}
\plotone{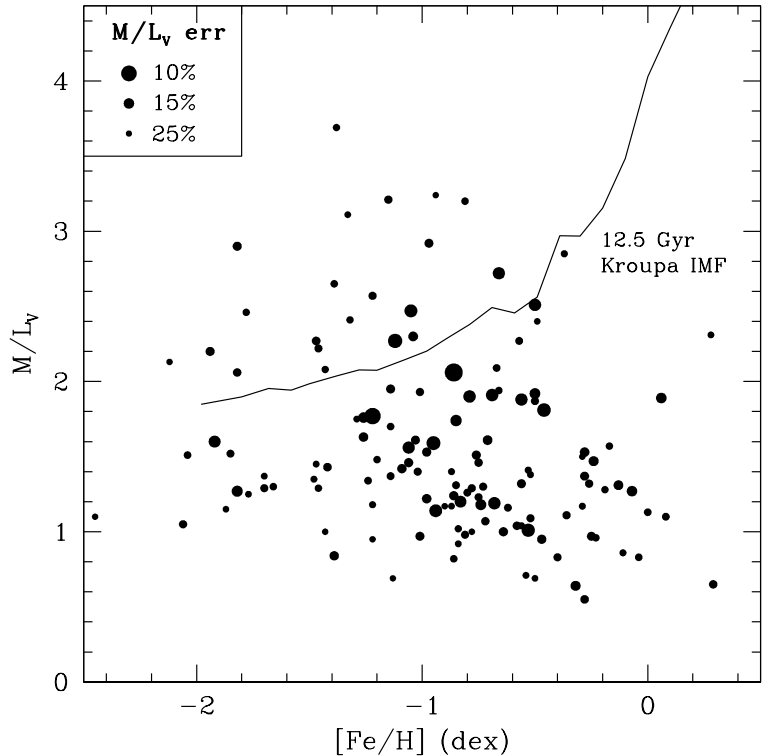}
\figcaption[z_good.eps]{\label{fig:fig_1}
$M/L_{V}$ vs.~metallicity for M31 GCs. The point size is inversely proportional to the $M/L_{V}$ error, as indicated in the legend. A single stellar population model curve for an age of 12.5 Gyr and a Kroupa initial mass function is overplotted (Conroy \& Gunn 2010). Metal-rich GCs deviate strongly from model predictions.}
\end{figure*}

Figure 2 shows the $K$-band mass-to-light ratios for M31 GCs. In the near-IR there is no similar line-blanketing effect expected and there is a minimal dependence of $M/L_K$ on metallicity. However, Figure 2 shows that $M/L_K$ strongly declines with increasing metallicity for M31 GCs. The apparent spread in the data is strikingly small, probably because the use of $K$ overcomes uncertainties in the reddening corrections. In addition, the relative offsets between the data and models in $V$ and $K$ help constrain the origin of the discrepancy, as we discuss  in \S 5.

\begin{figure*}
\epsscale{0.7}
\plotone{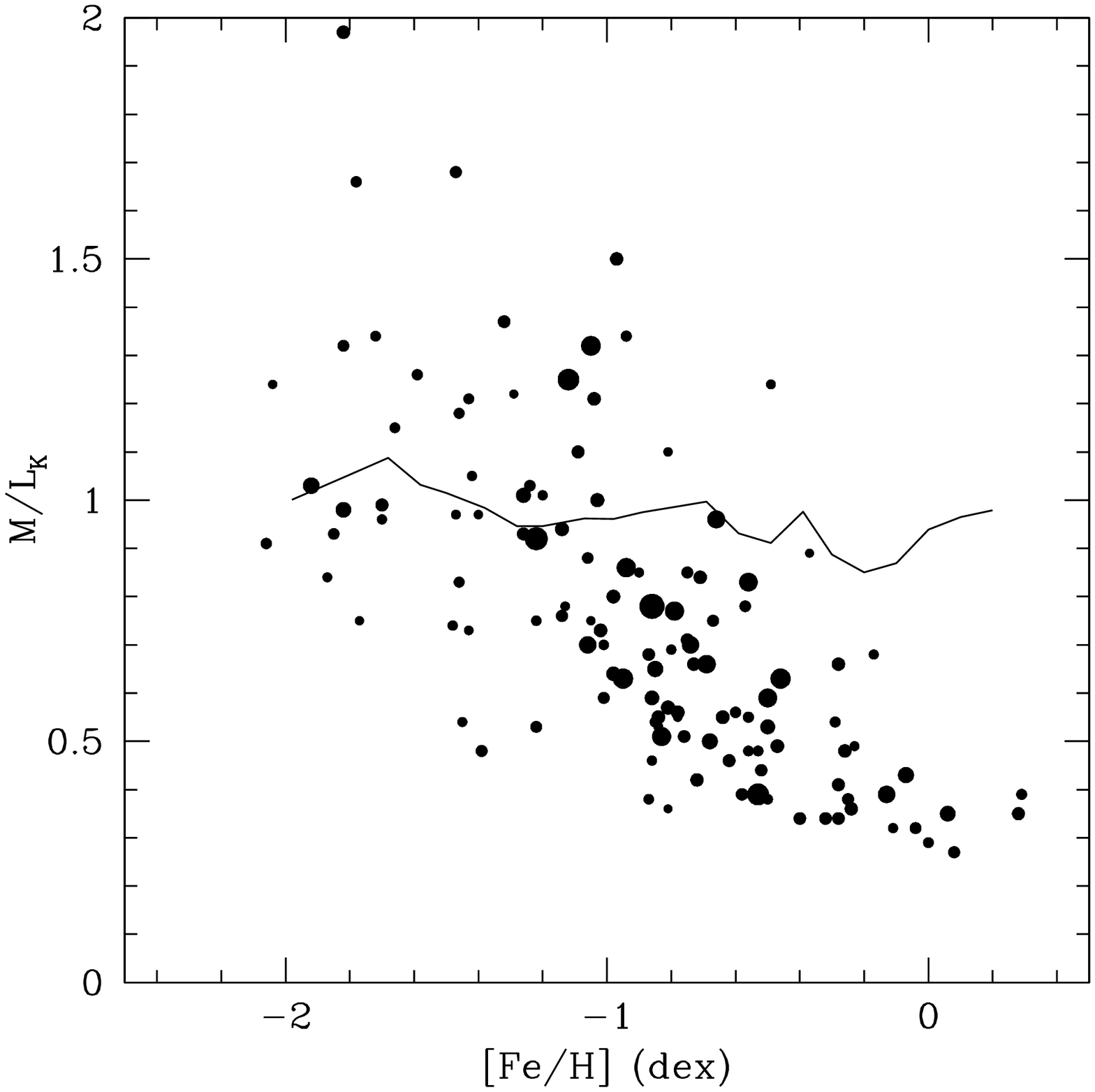}
\figcaption[z2_good.eps]{\label{fig:fig_2}
$M/L_{K}$ vs.~metallicity for M31 GCs. The point size is inversely proportional to the $M/L_{K}$ error, as indicated in the legend to Figure 1. A single stellar population model curve for an age of 12.5 Gyr and a Kroupa initial mass function is overplotted (Conroy \& Gunn 2010). $M/L_{K}$ for the M31 GCs decreases strongly with metallicity; the models predict no such trend.}
\end{figure*}

The M31 GC metallicities plotted in Figures 1 and 2 are calculated using an average of Lick indices that primarily measure iron. This average, in turn, is calibrated to the [Fe/H] scale of Galactic GCs using a two-part linear fit to Lick index measurements for 41 Galactic GCs in the range $-2.5 \la $ [Fe/H] $\la 0$ (see Caldwell \etal~2011 for details). Metallicities above solar are therefore extrapolations, and systematic offsets of order $\sim 0.1$--0.2 dex in the overall metallicity scale are possible, but these caveats do not impact any of our conclusions.

\subsection{A Reassuring Mass Trend}

This new sample of GCs is large enough that we can begin to explore internal dynamics as a function of cluster mass. Figure 3 plots $M/L_V$ against mass, with the GCs separated into two metallicity subpopulations at [Fe/H] = $-1$. The massive GCs ($\ga 10^{6} M_{\odot}$) have a median $M/L_V \sim 2$, while $M/L_V$ falls by nearly a factor of two near typical GC masses of 2--3 $\times 10^{5} M_{\odot}$. The scatter of a few GCs upward to high values of $M/L_V$ at intermediate masses may well be spurious; these are among the most uncertain data plotted.

A decline in $M/L_V$ toward lower masses is a basic theoretical expectation of dynamical evolution (e.g., Kruijssen 2008). As GCs evolve toward an isothermal state, driven largely by two-body relaxation, low-mass (high $M/L$) stars are preferentially ejected from the cluster, lowering its overall $M/L$. In Figure 3 we also plot two theoretical models of GC evolution from Kruijssen (2009, see also Kruijssen 2008). These models are for an intermediate metallicity of $-1.3$ and an age of 12.4 Gyr. In terms of the long term evolution, the metallicity assumed amounts largely to a zeropoint shift, and higher metallicity models do not fit the M31 data as discussed above, so we plot only a single metallicity. In these models, the mass loss due to dynamical evolution is assumed to be inversely proportional to a normalized dissolution timescale $t_0$; we plot models for $t_0 = 1$ and 3 Myr (these values correspond to total disruption timescales of approximately 14 and 43 Gyr, respectively, for a $10^{6} M_{\odot}$ cluster). Scatter is expected at all masses due to real variations in $t_0$ for each cluster, which depends on its structure and tidal history. The main locus of the GCs, both in terms of the ``break mass" and the slope at lower masses, appears more consistent with $t_0 = 1$ Myr, although the normalization is high below $\sim 10^{6} M_{\odot}$. Kruijssen also calculates models for an even lower $t_0$ (0.3 Myr), but these model GCs lose so many stars that their late $M/L$ evolution is dominated by dark remnants; hence, $M/L$ increases for the lowest mass GCs rather than decreasing. This predicted trend is inconsistent with the data in Figure 3 and we believe the $t_0 = 1$ Myr model best agrees with our observations.

\begin{figure*}
\epsscale{1.0}
\plottwo{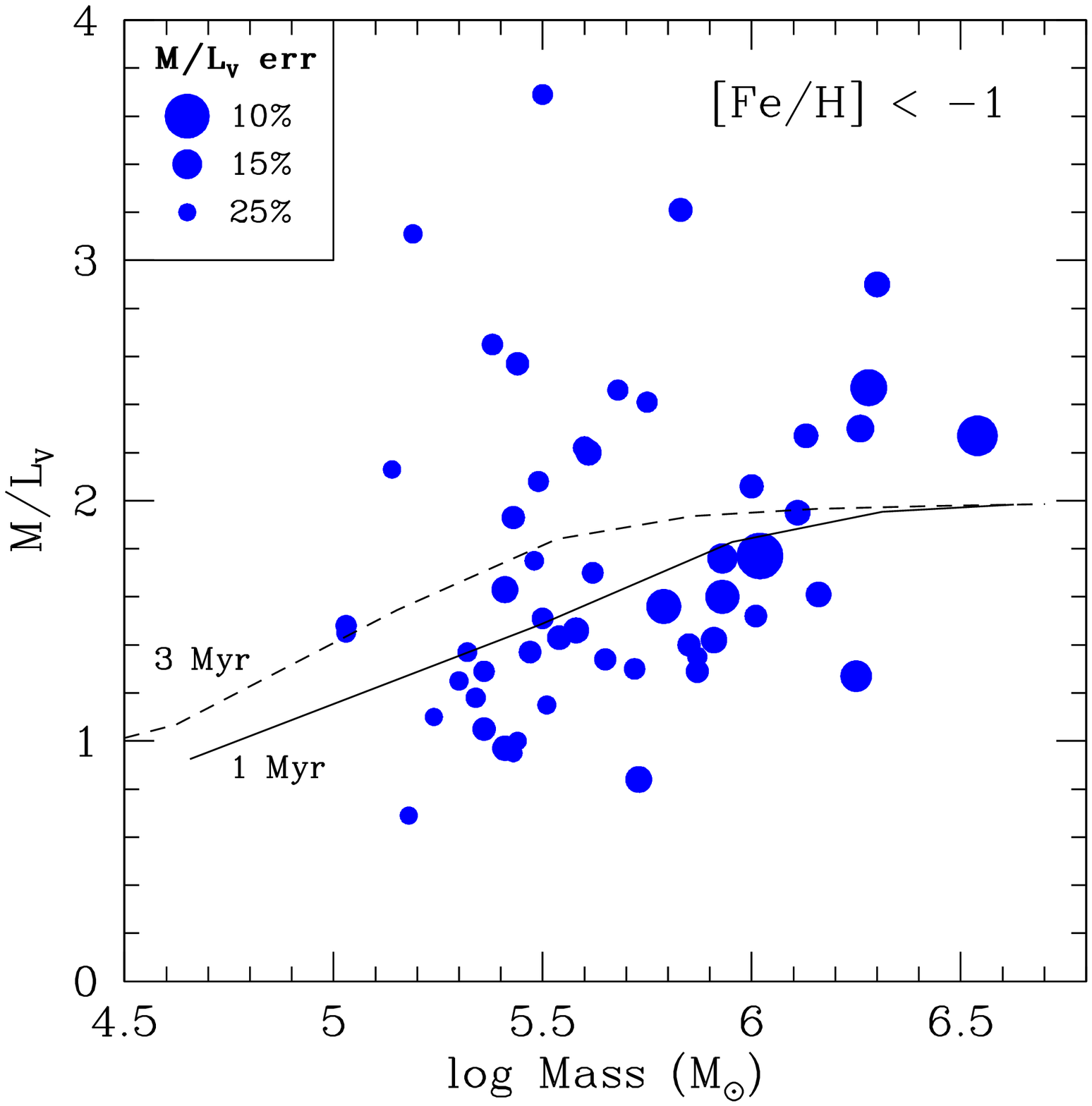}{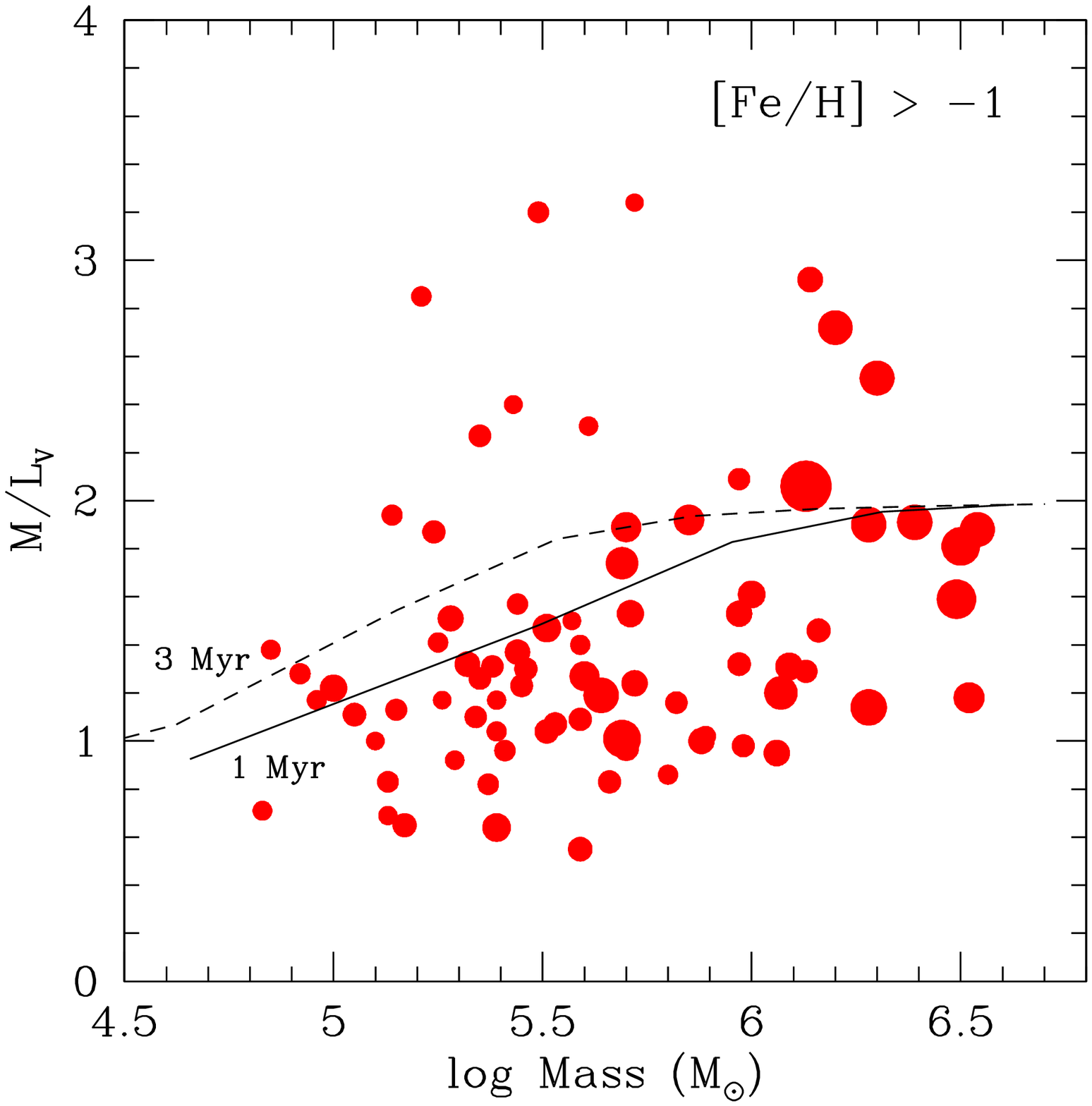}
\figcaption[f3.eps]{\label{fig:fig_3}
$M/L_{V}$ vs.~mass for M31 GCs, divided into metal-poor (left panel) and metal-rich (right panel) GCs. The point size is inversely proportional to the $M/L_{V}$ error, as indicated in the legend in the left panel. Model curves of cluster dynamical evolution from Kruijssen (2009), with dissolution timescales of 1 and 3 Myr, are plotted. The main locus of GCs for both subpopulations is consistent with a decrease in $M/L_{V}$ toward lower masses, as predicted by the models. Both subpopulations show similar trends, suggesting that dynamical evolution is not responsible for the discrepancy between the observed and predicted $M/L$ for metal-rich GCs.}
\end{figure*}

Despite the mismatch between the models and the data in detail, the general trends predicted are clear in our M31 data---and, indeed, much more so than in the Milky Way (e.g., Kruijssen \& Mieske 2009), due to the larger number of massive GCs observable in M31. 

Both metallicity subpopulations of GCs show similar trends in $M/L_V$ as a function of cluster mass. This shows that the discrepancy between the data and stellar population models for the metal-rich GCs is unlikely to have been caused by standard dynamical evolution.

\begin{figure*}
\epsscale{1.0}
\plottwo{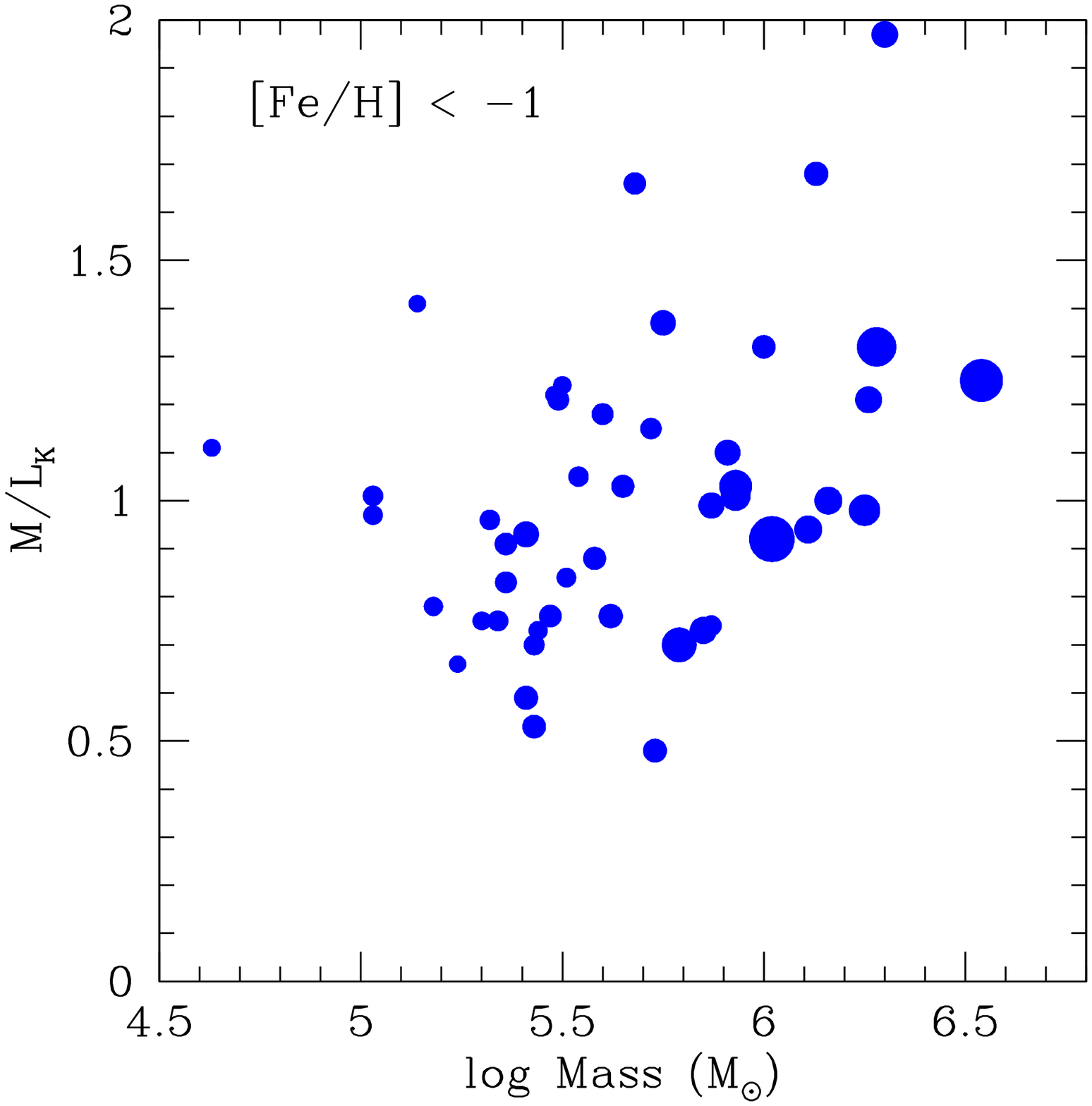}{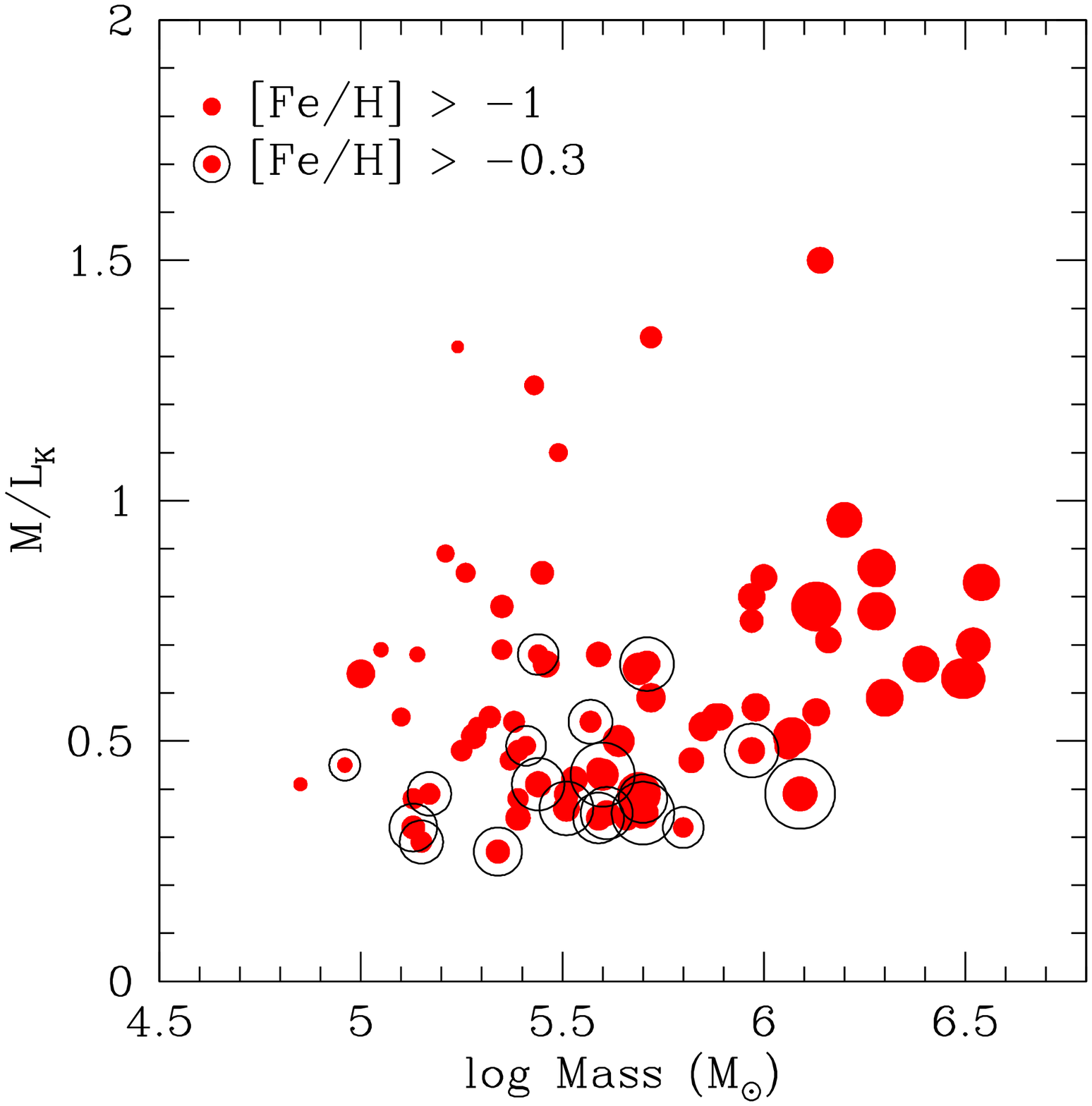}
\figcaption[f4.eps]{\label{fig:fig_4}
$M/L_{K}$ vs.~mass for M31 GCs, divided into metal-poor (left panel) and metal-rich (right panel) GCs. The point size is inversely proportional to the $M/L_{V}$ error, as indicated in Figure 3. The most metal-rich GCs are marked in the right panel, and have a surprisingly small range in $M/L_K$; dynamical evolution is unlikely to be the cause of these low estimates. Figure 2 shows that the expected values for GCs without a significant loss of low-mass stars
are $M/L_K \sim 0.95$, nearly independent of metallicity.}
\end{figure*}

Figure 4 plots $M/L_K$ vs.~cluster mass, again separated by subpopulations. The most metal-rich GCs (those with [Fe/H] $> -0.3$) are marked in the right panel. These objects have a small range in $M/L_K$ even over a large range in mass, and have lower $M/L_K$ at fixed mass than the metal-rich GCs of lower metallicity. This suggests that the ``extreme" low values of $M/L_K$ for the most metal-rich M31 GCs are probably not due to standard dynamical evolution, either, although this may be a mitigating effect.

It is worth considering whether mass is the most relevant dynamical variable for our comparisons. The evolution of an isolated GC is governed by its relaxation (thermalization) time ($t_{rh} \propto r^{3/2} \, M^{1/2}$) rather than its total mass. However, for GCs in a tidal field, recent theoretical works have emphasized that the most relevant dynamical variables are the mass and Jacboi radius (the theoretical tidal radius) and not the relaxation time (Baumgardt \& Makino 2003; Gieles \& Baumgardt 2008). The basic reason is that the fraction of stars lost per $t_{rh}$ also scales with cluster radius, essentially due to the finite size of the GC. As we mention above, even at fixed cluster mass, one expects variance in the degree of dynamical evolution. This is due to differing GC orbits and the resulting differences in tidal forces over the history of the cluster.

\begin{figure*}
\epsscale{0.55}
\plotone{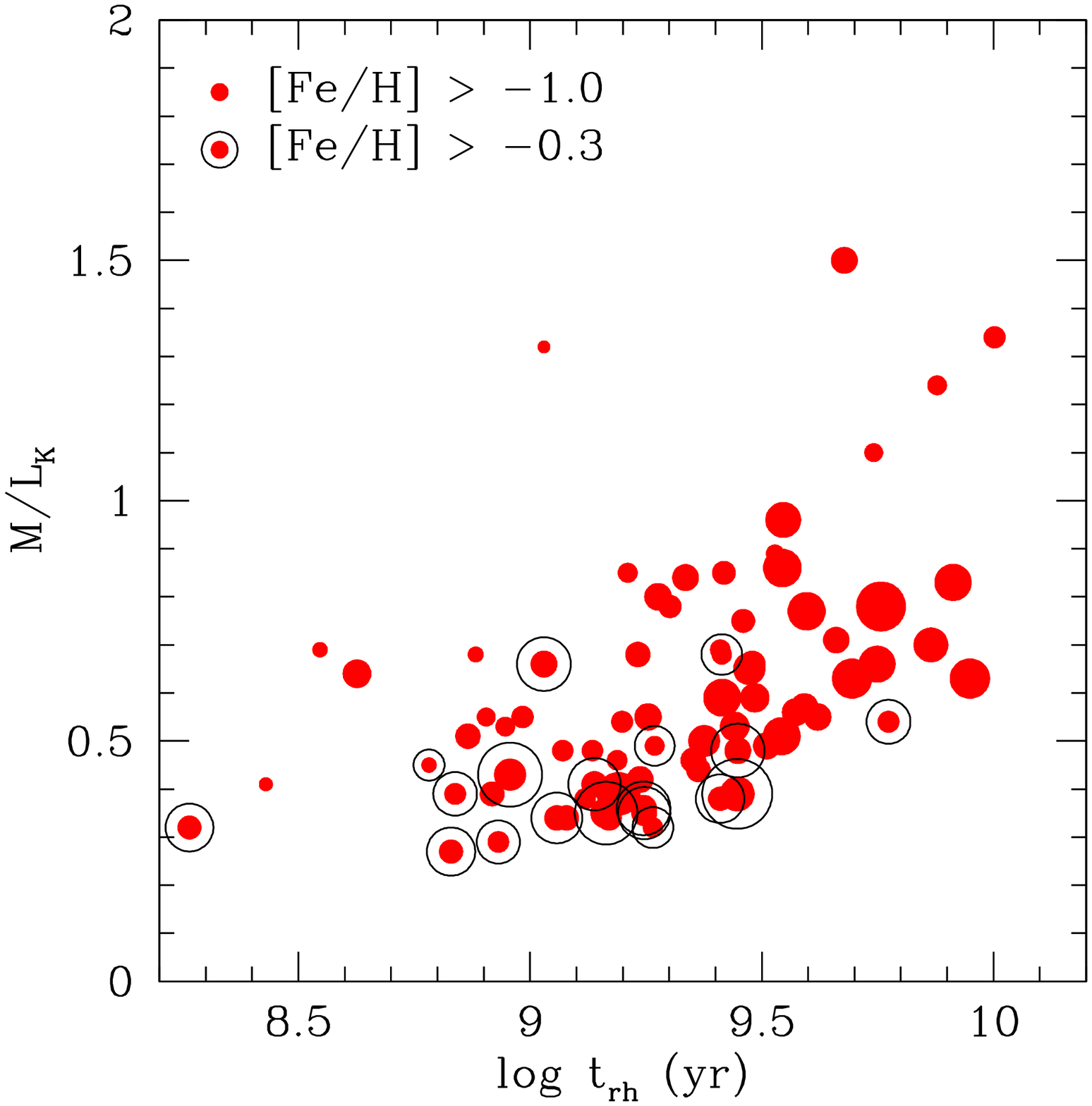}
\figcaption[f5.eps]{\label{fig:fig_5}
$M/L_{K}$ vs.~relaxation time ($t_{rh}$) for metal-rich M31 GCs. The point size is inversely proportional to the $M/L_{V}$ error, as indicated in Figure 3. The dispersion at fixed $t_{rh}$ is similar to that at fixed mass in Figure 4, so we cannot conclude whether mass or $t_{rh}$ is a more important determinant of dynamical evolution for M31 GCs.}
\end{figure*}

In Figure 5 we reproduce the right panel of Figure 4, but with relaxation time on the abscissa rather than mass. $t_{rh}$ is calculated using the equation in Spitzer (1987) and assuming a mean stellar mass of 0.3 $M_{\odot}$, appropriate for a Kroupa IMF. The scatter in $M/L_K$ at fixed $t_{rh}$ is perhaps slightly less than that at fixed mass in Figure 4, although interpreting these figures in strictly dynamical terms is not straightforward given the possibility of natal $M/L_K$ variations with metallicity. There is no strong evidence from these data for a preference between cluster mass and $t_{rh}$ as the dominant variable for dynamical evolution. Since only projected galactocentric radii are available for M31 GCs, it is challenging to calculate their expected evolution in a realistic tidal field (but see \S 4.3). However, in the near future, it will be possible to couple kinematic (radial velocity) data to improved structural parameters for many of these objects, enabling more detailed estimates of the expected evolution of individual GCs. Our $M/L$ dataset will then be among the best available to address the fundamental parameters of dynamical evolution.

\subsection{Confounding Factors and Caveats}

Here we briefly discuss additional physical factors that could influence our results or their interpretation.

There is some evidence among Galactic GCs that those with a deficit of low-mass stars, such as NGC 6712, have also suffered a larger than typical degree of external influence in the form of tidal shocking (e.g., de Marchi \etal~1999). Since metal-rich GCs are typically located at smaller galactocentric radii than metal-poor GCs, it seems plausible that metal-rich GCs have undergone accelerated dynamical evolution due to more pronounced tidal effects. Figure 6 shows $M/L_V$ as a function of projected galactocentric radius for both metal-poor and metal-rich GCs. While the mean $M/L_V$ is lower for the latter subpopulation (as discussed above), there is no evidence for a correlation for either subpopulation. Thus, tidal shocking related primarily to galactocentric radius does not appear to be an important driver of our results.

\begin{figure*}
\epsscale{1.0}
\plottwo{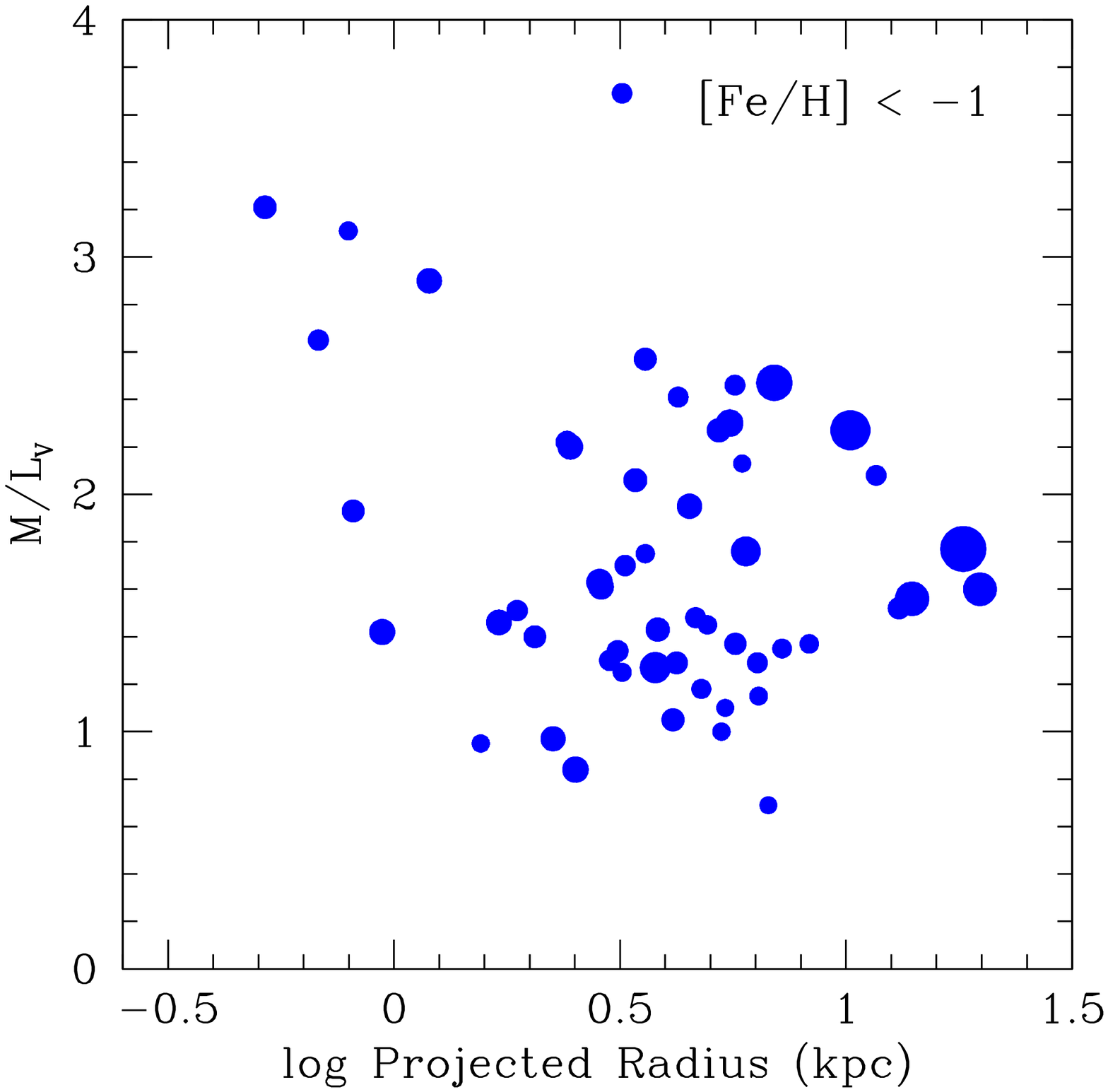}{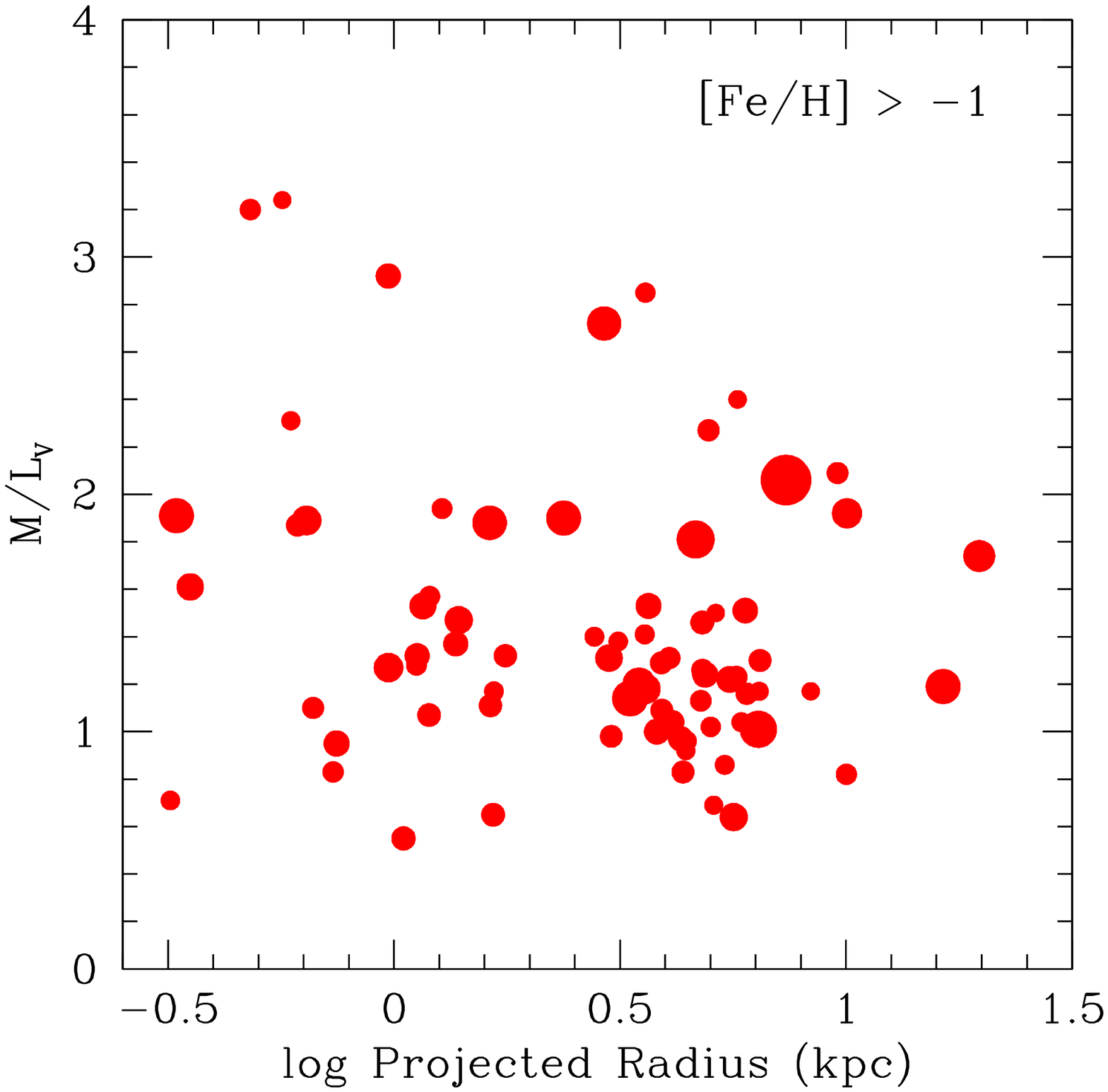}
\figcaption[col2.eps]{\label{fig:fig_6}
$M/L_{V}$ vs.~ projected galactocentric radius for M31 GCs, divided into metal-poor (left panel) and metal-rich (right panel) GCs.  The point size is inversely proportional to the $M/L_{V}$ error, as indicated in the legend to Figure 3. There is no evidence for a trend in either subpopulation, suggesting that accelerated dynamical evolution due to increased disk or bulge shocking at small galactocentric radii is probably not important for GCs in our sample.}
\end{figure*}

Mass segregation---in which heavier stars sink to the center of the GC---can lead to a bias, such that the measured velocity dispersion in a given aperture is larger than the equilibrium value for a single-mass King model. This will tend to inflate the mass estimate and thus $M/L$. This effect is the opposite of that observed in metal-rich M31 GCs, and, to the extent that it is relevant, will only increase the tension between the data and models.

Finally, we note that for GCs in the lower end of our mass range, stochastic effects can introduce additional scatter in $M/L$ at fixed mass, since a significant fraction of the cluster luminosity can be produced by a relatively small number of bright giants. This is especially relevant in the $K$ band as compared to $V$. We ran Monte Carlo simulations for a $10^{5} M_{\odot}$ cluster, with a Kroupa IMF and [Fe/H] = $-0.4$, utilizing Padova isochrones as above. These simulations suggest an additional spread ($1 \sigma$) of $\sim 4$\% in $M/L_V$ and $\sim 15$\% in $M/L_K$. If a subset of these GCs are depleted in low-mass stars, the actual effect may be still larger. Such stochastic variations are negligible for more massive GCs, so do not change our basic conclusions regarding trends of $M/L$ with metallicity and mass, but may contribute to the scatter (especially in $K$) for low-mass GCs.

\section{Discussion and Conclusions}

Our data appear to be most consistent with a scenario in which the initial optical and near-IR $M/L$ of M31 GCs depend on metallicity in a manner not reproduced by stellar population models. Subsequent dynamical evolution leads to the preferential loss of low-mass stars and a reduction in $M/L$,  and theoretical models of this process reproduce, at least in broad strokes, the basic features of the data. 

Previous studies of GCs in the Milky Way (e.g., Djorgovski 1995; McLaughlin 2000), dominated by metal-poor GCs, concluded that (at least in the core) they have an essentially constant $M/L_V \sim 2$. What has not been stressed before is that this result is \emph{not} consistent with stellar population models if metal-rich GCs are included in the sample. This can be seen more clearly in the $K$-band $M/L$ values of Figure 2, in which a strong anti-correlation between $M/L_K$ and metallicity is observed (see also Figure 3 of Djorgovski \etal~1997 for preliminary evidence of this same trend.)

In Strader \etal~(2009), we discussed several possible explanations for the lower than expected $M/L$ for metal-rich GCs. In essence, lowering $M/L$ with respect to a standard stellar population model requires an excess of stars with low $M/L$ (luminous giants, including red giant branch [RGB] or asymptotic giant branch [AGB] stars) or a deficit of stars with high $M/L$ (late K or M dwarfs). We consider these in turn.

\subsection{An Excess of Giants?}

We explore the first possibility using similar FSPS models to those discussed above\footnote{The BaSTI isochrones are now used instead of the Padova isochrones, since the latter do not explicitly separate stars by evolution phase.}, with a Kroupa IMF, an age of 12.5 Gyr, and a typical metal-rich GC metallicity of [Fe/H] = $-0.4$. For these parameters, the giants (red horizontal branch, RGB, and AGB) make up $\sim52$\% of the $V$ light and $\sim77$\% of the $K$ light. In the mean, the observed $M/L$ are low by a factor of $\sim 2$ in $V$ and 2--2.5 in $K$. Thus, it would be necessary to boost the amount of light from the giants by a factor of 2--3 at fixed mass to match the models. Schiavon \etal~(2002) compare observed and model luminosity functions on the upper RGB of the metal-rich GC 47 Tucanae and argue that the models may underpredict the number of upper giant branch stars by 50--60\%. This could lead to an overall discrepancy of perhaps 25\% in giant light (since the offset is observed only on the upper giant branch), which is significant but not sufficient to explain our $M/L$ observations. In addition, similar offsets have been reported in some metal-poor GCs (e.g., Langer \etal~2000), so their relevance to this problem is uncertain. 

Considering the AGB, Girardi \etal~(2010) show that the Padova models overpredict the lifetimes of the most luminous AGB stars for old metal-poor stellar populations. While this finding obviously does not directly affect the metal-rich GCs, it could still be germane to solving our problem. Correcting this issue will increase the predicted $M/L$ for metal-poor GCs (since the models will have fewer low $M/L$ stars), which could lead to higher than expected $M/L$ for both the metal-poor and metal-rich GCs. This idea, combined with a metallicity-independent loss of low-mass stars (through, for example, an unexpected dynamical mechanism), could then explain the full set of observations. This has some appeal, since a metallicity-dependent IMF or dynamical process might otherwise be required (see \S5.2).

We conclude that it is unlikely that an excess of bright giants is the sole or even principal reason for the unusual $M/L$ observed among metal-rich M31 GCs, although this problem deserves further attention.

\subsection{A Deficit of Dwarfs?}

The second possibility---a metallicity-dependent deficit of low-mass dwarfs---could be either primordial (from a``bottom-light" IMF) or due to the preferential loss of such stars through dynamical evolution. As discussed in \S 4, and confirming our results from Strader \etal~(2009), this latter scenario does not seem feasible. Even though we see clear evidence for such evolution among M31 GCs of lower mass, the discrepancy between the data and models persists for the most massive M31 GCs, and these objects are expected to have undergone little evolution. Additional dynamical mass measurements for massive GCs, especially those at larger projected galactocentric radii, would be useful.

Next we consider what IMF can reproduce the optical and near-IR $M/L$ values for metal-rich GCs. We again use the Conroy \& Gunn (2010) FSPS models, with an initial mass range between 0.1 and 100 $M_{\odot}$. For an IMF of the form $dN/dM \propto M^{-\alpha}$, the default Kroupa-style IMF has three mass ranges: $< 0.5 M_{\odot}$, between 0.5 and 1.0 $M_{\odot}$, and $> 1.0 M_{\odot}$, which we define with indices $\alpha_1$, $\alpha_2$, and $\alpha_3$ respectively. We hold $\alpha_3 = 2.3$ in all cases ($\alpha_3$ affects only the stellar remnants for the old ages of GCs). The default values are $\alpha_1 = 1.3$ and $\alpha_2 = 2.3$ (for a Salpeter IMF, one would have $\alpha$ = 2.3 for all masses). We have calculated models, again at a reference metallicity of $-0.4$, for a wide range of ($\alpha_1$, $\alpha_2$) pairs in steps of 0.5. These pairs are all shallower (and thus with fewer low-mass dwarfs) than the default values.

\begin{figure*}
\epsscale{0.7}
\plotone{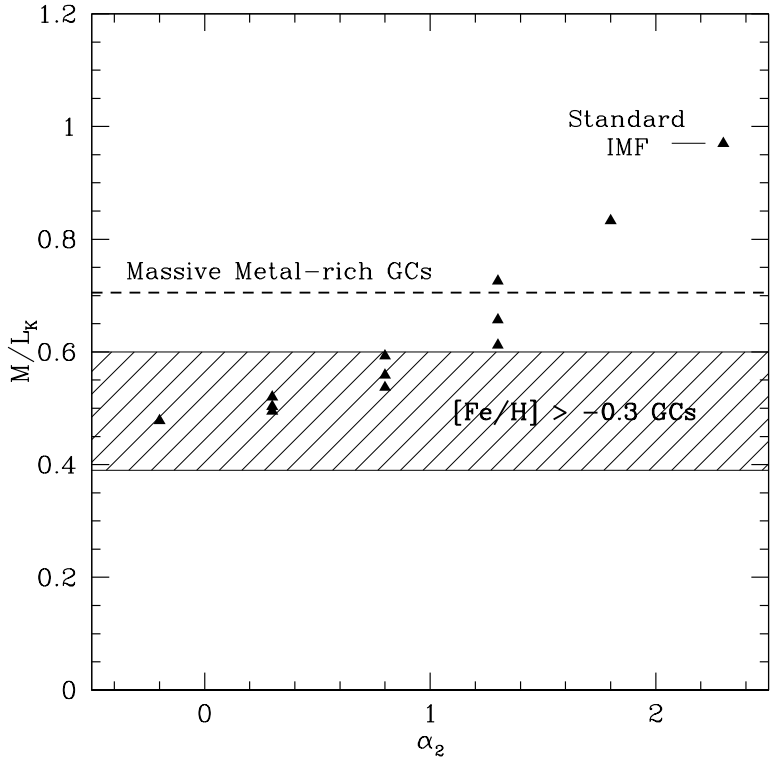}
\figcaption[f7.eps]{\label{fig:fig_7}
$M/L_{K}$ vs.~ $\alpha_2$ (the slope of the mass function in the range $0.5 < M < 1.0 \,M_{\odot}$) for M31 GCs. The points are model predictions at a fixed metallicity of [Fe/H] = $-0.4$ but for varying initial mass function slopes $\alpha_2$ and $\alpha_1$ (the slope below $0.5 \,M_{\odot}$). The dashed line is the median observed $M/L_K$ for massive metal-rich GCs. The shaded region covers a range of assumptions for the most metal-rich GCs ([Fe/H] $> -0.3$), as described in \S 5.2. A relatively shallow mass function below $1 \, M_{\odot}$, with a slope from $\sim 0.8 - 1.3$, is favored. The model corresponding to a  standard Kroupa IMF is labeled. The exact ($\alpha_1$, $\alpha_2$) pairs listed are, from highest to lowest $M/L_K$ for each value of $\alpha_2$: (1.3, 2.3), (1.3, 1.8), (1.3, 1.3), (0.8, 1.3), (0.3, 1.3), (0.8, 0.8), (0.3, 0.8), ($-0.2, 0.8$), (0.3, 0.3), ($-0.2, 0.3$), ($-0.5, 0.3$), ($-0.2, -0.2$).}
\end{figure*}

We plot the resulting $M/L_{K}$ predictions in Figure 7 as a function of $\alpha_2$. The upper dotted line is the median $M/L_{K}$ for massive metal-rich GCs---those expected to have undergone little dynamical evolution. This value is $M/L_{K} = 0.71$, consistent with both $\alpha_1$ and $\alpha_2 \sim 1.3$. The lower shaded region is for the most metal-rich M31 GCs ([Fe/H] $> -0.3$). Few of these GCs have masses $> 10^6 M_{\odot}$ and most are expected to have undergone dynamical evolution that reduced $M/L_{K}$. The upper boundary of the shaded region, $M/L_{K} = 0.6$, is an approximate ``unevolved" value for $M/L_{K}$ of the most metal-rich GCs, assuming the Kruijssen (2009) models with $t_0 = 1$ Myr. In other words, this is the 
$M/L_{K}$ expected if the models were correct and if we had massive, minimally evolved GCs in the subsample. The lower boundary of the shaded region is the observed median of the [Fe/H] $> -0.3$ sample, $M/L_{K} = 0.39$. This value is lower than that predicted by the models for any IMF, implying that these GCs must have lost significant numbers of low-mass stars through dynamical evolution. The upper boundary of the shaded region is most consistent with the models ($\alpha_1$, $\alpha_2$)  = (0.3, 1.3) and (0.8, 0.8). These models predict $M/L_V \sim 1.9$, which is reasonably consistent with our typical estimates for massive metal-rich GCs, although there is a large spread in $M/L_V$ at fixed metallicity. While not plotted, more extreme combinations are also consistent with the data, e.g., a normal slope of $\alpha_2 = 2.3$ but the removal of all stars with masses $< 0.5 M_{\odot}$. Such models are less plausible, but are not ruled out by the $M/L$ data alone. 

Since the predicted $M/L_K$ varies in an essentially monotonic fashion with the slope of the mass function, and there is minimal dependence on metallicity, Figure 7 can be used to interpret observations at lower metallicities. Of course, at these metallicities, the deviations from stellar population model predictions are less substantial, and some of the other effects discussed earlier could be dominant.

We conclude that a shallow mass function below 1 $M_{\odot}$, of the approximate form $dN/dM \propto M^{-0.8} -  M^{-1.3}$, is a viable explanation for the unusual optical and near-IR $M/L$ for metal-rich M31 GCs. By the arguments outlined above, we believe that such a mass function is unlikely to be caused by standard dynamical evolution; it may reflect either the initial stellar mass function in these GCs, or the loss of low-mass stars (probably early in the cluster's life) through other means. One possibility is suggested by Marks \& Kroupa (2010), who propose metallicity-dependent cluster winds, with resulting expansion, as an explanation for the preferential loss of low-mass stars in more metal-rich GCs. 

As discussed in Strader \etal~(2009), the minimal available evidence suggests that metal-rich Galactic GCs follow similar $M/L$ trends as in M31. Therefore, Galactic GCs may offer a convenient avenue for studying this phenomenon in more depth, including detailed dynamical studies and direct estimates of the mass function.

\subsection{The Mass Function in Context}

As discussed in Strader \etal~(2009), there is a long history of suggestions that the stellar mass function of Galactic GCs varies with metallicity, such that more metal-rich GCs have flatter present-day mass functions. This issue has most recently been considered by Paust \etal~(2010), who present measurements for the slope of the mass function in 17 Galactic GCs using HST/ACS data. While they argue there is no evidence for a trend with metallicity in their sample, only two of their GCs have [Fe/H] $> -1$. Giersz \& Heggie (2011), in Monte Carlo modeling of 47 Tuc, find a best-fit IMF slope of $\sim 0.4$ below 0.8 $M_{\odot}$; this mass function is substantially flatter than in a Kroupa IMF and is more consistent with our M31 results.

van Dokkum \& Conroy (2010) used gravity-sensitive features in the spectra of Virgo and Coma giant elliptical galaxies to argue that these galaxies have a very bottom-heavy stellar mass function below $1 M_{\odot}$, with a slope of $\ga 3$ (even steeper than Salpeter). Other recent studies of early-type galaxies, using a variety of dynamical methods, have also argued for the presence of more low-mass stars than predicted from a Kroupa IMF (Treu \etal~2010; Thomas \etal~2011), although this conclusion is not universal (Cappellari \etal~2006). 
These galaxies have approximately solar metallicity and so are comparable to the most metal-rich M31 GCs. Our observations do not provide support for these claims of steep mass functions: assuming they are correct, either the IMF in these GCs is different than in massive elliptical galaxies, or the GCs have lost an enormous fraction of their low-mass stars (a much larger fraction than we discuss above for a Kroupa IMF, which we already found difficult to explain with standard dynamical evolution). A straightforward resolution is not clear, but the interface of star cluster and galaxy research on the stellar mass function promises to be fruitful.

\acknowledgments

We thank Michele Cappellari, Charlie Conroy, Aaron Dotter, Diederick Kruijssen, Soeren Larsen, Steinn Sigurdsson, and Graeme Smith for useful comments and conversations, Susan Tokarz for help with the spectroscopic data reduction, and Perry Berlind and Michael Calkins for help acquiring the data. We thank the anonymous referee for a useful report. This work was initiated while J.~S.~was supported by NASA through a Hubble Fellowship, administered by the Space Telescope Science Institude, which is operated by the Association of Universities for Research in Astronomy, Incorporated, under NASA contract NAS5-26555. Some of the observations reported here were obtained at the MMT Observatory, a joint facility of the Smithsonian Institution and the University of Arizona. This paper uses data products produced by the OIR Telescope Data Center, supported by the Smithsonian Astrophysical Observatory. Based on observations made with the NASA/ESA Hubble Space Telescope, and obtained from the Hubble Legacy Archive, which is a collaboration between the Space Telescope Science Institute (STScI/NASA), the Space Telescope European Coordinating Facility (ST-ECF/ESA) and the Canadian Astronomy Data Centre (CADC/NRC/CSA). This work was partially supported by the National Science Foundation through grant AST-0808099.


\LongTables
\newpage

\begin{deluxetable}{llrrr}
\tablewidth{0pt}
\tablecaption{Observing Log
        \label{tab:obslog}}
\tablehead{Order & Date & Central $\lambda$ & Total Exp.~Time & Seeing \\
           &  & (\AA) &  (hr) & (arcsec) }

\startdata

RV31 & 2008 Oct 23       &   5215   & 3.0 &  1.7  \\
RV31 & 2009 Oct 24       &    5215  & 2.3 &  0.4 \\
Ca19 & 2008 Nov 08      &  8542   & 3.0 &  0.8  \\
Ca19 & 2008 Nov 09      &   8542  & 3.0 &  0.6  \\
OB37 & 2008 Oct 24/25 &   4340  & 5.0 &  0.4  \\
OB25 & 2008 Oct 21       &  6540   & 3.0  &  1.0  \\

\enddata

\end{deluxetable}

\begin{deluxetable}{lrrl}
\tablewidth{0pt}
\tablecaption{Heliocentric Radial Velocities
        \label{tab:rv}}
\tablehead{ID & R.~V. & error & age\tablenotemark{a} \\
           &  \kms &  \kms  & }

\startdata

AU010 & $-324.9$ & 0.6 & old \\
B003-G045 & $-381.4$ & 0.6 & old \\
B005-G052 & $-283.4$ & 0.6 & old \\
B006-G058 & $-236.0$ & 0.4 & old \\
B008-G060 & $-323.2$ & 0.4 & old \\
B010-G062 & $-161.7$ & 0.6 & old \\
B010D & $-368.5$ & 2.0 & young \\
B012-G064 & $-360.7$ & 0.6 & old \\
\enddata

\tablenotetext{a}{Categories are: old ($> 3$ Gyr), young ($< 1$ Gyr), inter (between 1 and 3 Gyr) and unk (unknown age).}

\end{deluxetable}

\begin{deluxetable}{lrrrrr}
\tablewidth{0pt}
\tablecaption{Individual Order Velocity Dispersion Estimates
        \label{tab:sig_raw}}
\tablehead{ID & RV31 (2008) & RV31 (2009) & OB37 & Ca19 & Avg. \\
           &  \kms &  \kms  &  \kms &  \kms &  \kms }

\startdata

AU010  &  \nodata  &  $7.0\pm0.4$  &  \nodata  &  \nodata  &  $7.0\pm0.6$  \\
B003-G045  &  $6.6\pm0.6$  &  \nodata  &  $7.2\pm1.0$  &  \nodata  &  $6.7\pm0.6$  \\
B005-G052  &  \nodata  &  $15.6\pm0.5$  &  \nodata  &  \nodata  &  $15.6\pm0.9$  \\
B006-G058  &  $12.1\pm0.5$  &  \nodata  &  $11.7\pm0.6$  &  \nodata  &  $12.0\pm0.7$  \\
B008-G060  &  $7.4\pm0.5$  &  \nodata  &  $6.8\pm0.5$  &  \nodata  &  $7.2\pm0.5$  \\
B010-G062  &  $5.8\pm0.5$  &  \nodata  &  $7.6\pm0.6$  &  \nodata  &  $6.6\pm0.5$  \\
B012-G064  &  $19.9\pm0.5$  &  \nodata  &  $18.3\pm0.7$  &  \nodata  &  $19.3\pm1.0$  \\
B013-G065  &  $5.4\pm0.5$  &  \nodata  &  $3.6\pm0.9$  &  \nodata  &  $5.0\pm0.5$  \\
\enddata

\end{deluxetable}

\begin{deluxetable}{lrrrrrrrrr}
\tablewidth{0pt}
\tablecaption{New Dynamical Quantities
        \label{tab:new_dyn}}
\tablehead{ID & $\sigma_{0}$  & $\sigma_{\infty}$ & log $M_{vir}$ & log $M_{King}$ & log $L_V$ & log $L_K$ & $M/L_{V}$  & $M/L_{K}$  & [Fe/H] \\  
                         &  (\kms)               &   (\kms)                  & ($M_{\odot}$) & ($M_{\odot}$) & ($L_{\odot}$) & ($L_{\odot}$) &  & & (dex) }

\startdata

AU010 & $8.0\pm0.6$ & $6.1\pm0.5$ & $5.24_{-0.09}^{+0.07}$ & $5.17_{-0.12}^{+0.09}$ & $4.97\pm0.01$ & $5.12\pm0.11$ & $1.87\pm0.36$ & $1.32\pm0.46$ & $-0.50$ \\
B005-G052 & $17.3\pm1.0$ & $13.3\pm0.8$ & $5.97_{-0.08}^{+0.07}$ & $5.91_{-0.09}^{+0.08}$ & $5.65\pm0.03$ & $6.09\pm0.01$ & $2.09\pm0.42$ & $0.75\pm0.14$ & $-0.67$ \\
B006-G058 & $13.1\pm0.8$ & $10.6\pm0.6$ & $5.69_{-0.06}^{+0.05}$ & $5.66_{-0.06}^{+0.06}$ & $5.69\pm0.02$ & $6.10\pm0.01$ & $1.00\pm0.14$ & $0.39\pm0.05$ & $-0.53$ \\
B008-G060 & $7.6\pm0.5$ & $6.1\pm0.4$ & $5.45_{-0.08}^{+0.07}$ & $5.42_{-0.11}^{+0.09}$ & $5.36\pm0.03$ & $5.52\pm0.02$ & $1.23\pm0.24$ & $0.85\pm0.16$ & $-0.75$ \\
B010-G062 & $6.7\pm0.5$ & $5.6\pm0.4$ & $5.68_{-0.09}^{+0.08}$ & $5.62_{-0.12}^{+0.09}$ & $5.29\pm0.03$ & $5.46\pm0.01$ & $2.46\pm0.52$ & $1.66\pm0.33$ & $-1.78$ \\
B012-G064 & $20.4\pm1.1$ & $16.1\pm0.9$ & $6.34_{-0.07}^{+0.06}$ & $6.29_{-0.07}^{+0.06}$ & $5.84\pm0.02$ & $6.03\pm0.01$ & $3.13\pm0.50$ & $2.03\pm0.32$ & $-1.66$ \\
B013-G065 & $5.0\pm0.5$ & $4.2\pm0.4$ & $5.43_{-0.11}^{+0.09}$ & $5.34_{-0.13}^{+0.10}$ & $5.05\pm0.03$ & $5.34\pm0.02$ & $2.40\pm0.57$ & $1.24\pm0.28$ & $-0.49$ \\
B017-G070 & $13.2\pm0.8$ & $10.4\pm0.6$ & $5.89_{-0.07}^{+0.06}$ & $5.85_{-0.08}^{+0.07}$ & $5.88\pm0.06$ & $6.15\pm0.01$ & $1.02\pm0.22$ & $0.55\pm0.09$ & $-0.84$ \\
\enddata

\end{deluxetable}

\begin{deluxetable}{lrrrrr}
\tablewidth{0pt}
\tablecaption{New King Model Structural Parameters
        \label{tab:new_king}}
\tablehead{ID & $c$ & $r_h$ & $r_0$ & Filter & Instrument \\
           &   &  (pc)  &  (pc) &   &  }

\startdata

AU010  &  1.99  &  2.05  &  0.26  &  F475W  &  ACS  \\
B008-G060  &  1.53  &  3.18  &  1.10  &  F435W  &  ACS  \\
B010-G062  &  1.24  &  6.66  &  3.42  &  F435W  &  ACS  \\
B013-G065  &  1.08  &  6.53  &  4.07  &  F814W  &  ACS  \\
B020D-G089  &  1.83  &  2.78  &  0.53  &  F814W  &  WFPC2  \\
B029-G090  &  1.67  &  4.99  &  1.35  &  50CCD  &  STIS  \\
B045D  &  1.67  &  1.18  &  0.32  &  F814W  &  ACS  \\
B056-G117  &  1.47  &  1.89  &  0.72  &  F435W  &  ACS  \\
\enddata

\tablecomments{The listed structural parameters are concentration $c$ = log($r_t/r_0$), half-light radius ($r_h$),
and King radius ($r_0$). The errors are assumed to be 15\% in $r_t/r_0$ and 10\% in the radii.}
\tablenotetext{a}{The core radius of the GC B159 is too small to measure; it is possible that this cluster has undergone core collapse.
B159 hosts an X-ray source, consistent with a high central density (Peacock \etal~2010).}

\end{deluxetable}

\begin{deluxetable}{lrrrrrrrrrl}
\tablewidth{0pt}
\tablecaption{Revised Literature Dynamical Quantities
        \label{tab:old_dyn}}
\tablehead{ID & $\sigma_{0}$  & $\sigma_{\infty}$ & log $M_{vir}$ & log $M_{King}$ & log $L_V$ & log $L_K$ & $M/L_{V}$  & $M/L_{K}$  & [Fe/H] & source \\  
                         &  (\kms)               &   (\kms)                  & ($M_{\odot}$) & ($M_{\odot}$) & ($L_{\odot}$) & ($L_{\odot}$) &  & & (dex) &  }

\startdata

B006-G058 & $13.2\pm0.6$ & $10.7\pm0.5$ & $5.69_{-0.04}^{+0.04}$ & $5.66_{-0.05}^{+0.05}$ & $5.69\pm0.02$ & $6.10\pm0.01$ & $1.01\pm0.12$ & $0.39\pm0.04$ & $-0.53$ & D97;DG97;S09 \\
B012-G064 & $17.9\pm2.4$ & $14.1\pm1.9$ & $6.22_{-0.15}^{+0.11}$ & $6.18_{-0.15}^{+0.11}$ & $5.84\pm0.02$ & $6.03\pm0.01$ & $2.41\pm0.71$ & $1.56\pm0.46$ & $-1.66$ & D97 \\
B020-G073 & $16.8\pm0.5$ & $13.6\pm0.4$ & $6.13_{-0.03}^{+0.03}$ & $6.11_{-0.03}^{+0.03}$ & $5.82\pm0.01$ & $6.24\pm0.01$ & $2.06\pm0.18$ & $0.78\pm0.07$ & $-0.86$ & D97;DG97 \\
B023-G078 & $27.7\pm5.1$ & $22.9\pm4.2$ & $6.63_{-0.20}^{+0.14}$ & $6.60_{-0.20}^{+0.14}$ & $6.48\pm0.02$ & $6.86\pm0.01$ & $1.42\pm0.53$ & $0.59\pm0.22$ & $-0.74$ & D97 \\
B034-G096 & $14.3\pm0.7$ & $11.4\pm0.5$ & $5.82_{-0.08}^{+0.07}$ & $5.79_{-0.09}^{+0.08}$ & $5.75\pm0.04$ & $6.15\pm0.02$ & $1.16\pm0.23$ & $0.46\pm0.08$ & $-0.62$ & S09 \\
B037-V327 & $21.2\pm2.6$ & $17.7\pm2.2$ & $6.47_{-0.13}^{+0.10}$ & $6.42_{-0.13}^{+0.10}$ & $6.20\pm0.24$ & $6.83\pm0.03$ & $1.91\pm1.49$ & $0.45\pm0.12$ & $-0.81$ & C06 \\
B045-G108 & $10.8\pm0.8$ & $8.9\pm0.7$ & $5.72_{-0.07}^{+0.06}$ & $5.69_{-0.08}^{+0.07}$ & $5.63\pm0.02$ & $5.95\pm0.01$ & $1.24\pm0.21$ & $0.59\pm0.09$ & $-0.86$ & D97;DG97;C09 \\
B058-G119 & $23.2\pm1.0$ & $18.8\pm0.8$ & $6.28_{-0.04}^{+0.04}$ & $6.25_{-0.05}^{+0.05}$ & $5.89\pm0.02$ & $6.16\pm0.02$ & $2.47\pm0.30$ & $1.32\pm0.15$ & $-1.05$ & S09 \\
\enddata

\tablecomments{The sources for the velocity dispersions are: D97---Djorgovski \etal~(1997); DG97---Dubath \& Grillmair (1997); C06---Cohen (2006); S09---Strader \etal~(2009); C09---Colucci \etal~(2009).}

\end{deluxetable}

\end{document}